\documentclass[fleqn,usenatbib]{mnras}

\usepackage{newtxtext,newtxmath}

\usepackage[T1]{fontenc}

\DeclareRobustCommand{\VAN}[3]{#2}
\let\VANthebibliography\thebibliography
\def\thebibliography{\DeclareRobustCommand{\VAN}[3]{##3}\VANthebibliography}

\usepackage{graphicx}	
\usepackage{amsmath}	
\usepackage{pdflscape}
\usepackage{multirow}

\title[]{Propeller states in locally super-critical ULXs}

\author[M. Middleton et al.]{
Middleton, M.$^{1}$, G\'{u}rpide, A$^{1}$ \& Walton, D. J$^{2}$,
\\
$^{1}$School of Physics \& Astronomy, University of Southampton, Southampton, Southampton SO17 1BJ, UK\\
$^{2}$ Centre for Astrophysics Research, University of Hertfordshire, College Lane, Hatfield AL10 9AB, UK
}

\date{Accepted XXX. Received YYY; in original form ZZZ}

\pubyear{2022}

\begin{document}
\label{firstpage}
\pagerange{\pageref{firstpage}--\pageref{lastpage}}
\maketitle

\begin{abstract}
An expected signature of the presence of neutron stars in the population of ultraluminous X-ray sources (ULXs) are large scale changes in X-ray luminosity, as systems reach spin equilibrium and a propeller state ensues. We explore the predicted luminosity changes when the disc is locally super-critical, finding that a significant parameter space in dipole field strength and accretion rate (at large radius) can be accompanied by changes of less than an order of magnitude in luminosity. We discuss the spectral signature and locate three ULXs (IC 342 X-1, Cir ULX-5 and NGC 1313 X-1) which appear to show changes consistent with super-Eddington systems entering a propeller state, and place rough constraints on the dipole field strength of NGC 1313 X-1 of $<$ 10$^{10}$ G. This work implies that the most reliable means by which to search for putative propeller states will be to search for changes in hardness ratio and at high energies.

\end{abstract}




\section{Introduction}

The population of ultraluminous X-ray sources (ULXs) is undoubtedly a mixture of stellar mass black holes and neutron stars, the vast majority of which are accreting at super-critical rates (the possibility remains that a small number of ULXs contain black holes of $\sim$100 M$_{\odot}$ or more, accreting at or below their Eddington limit). Determining the nature of the intrinsic and observed population is important as it relates to the process of super-critical accretion, such as the role of geometrical beaming (e.g. \citealt{King_2009_Beaming,Middleton_King2016, Middleton_2017_demographics_from_beaming, Wiktorowicz_2019_obs_vs_tot, Khan2022}) and the strength of neutron star magnetic dipole fields (e.g. \citealt{Mushtukov2015, King_2017_Pulsating_ULXs, Middleton_Brightman_2019_M51, King_Lasota_2020_PULX_iceberg_emerges, Brice2021, Kong2022_multipole}), and is needed for simulations relying on populations of accreting binaries (e.g. \citealt{Fragos_2013_reionization, Kuranov2020, Kovlakas2022}).

Locating neutron star ULXs (NS-ULXs hereafter) has relied on two approaches, already widely used by the X-ray binary community: searches for pulsations, and cyclotron resonance scattering features (CRSFs). The former have been detected in an increasing number of pulsing ULXs (PULXs, \citealt{NSULX_Bachetti_2014,Fuerst_P13_2021,Israel_2017_NSULX_5907, Carpano_2018_NSULX_NGC300, Sathyaprakash_2019_1313X2, 2019_Vasilopoulos_MNRAS.488.5225V,2020_Rodriguez_ApJ...895...60R}), whilst features resembling CRSFs have been located or inferred to be present in three ULXs to-date (\citealt{2018_Brightman_CRSF, Walton_2018_CRSF, Kong2022_multipole} although see also \citealt{Koliopanos_NGC300_CRSF}). It may be intrinsically very difficult to locate pulsations or CRSFs due to reprocessing in the cone of the wind or in the magnetospheric curtain if optically thick (\citealt{Mushtukov_2017_opt_thick_env_NS, 2021_Mushtukov_MNRAS.501.2424M}), due to a lack of sufficient data quality, or simply due to the observing band being too narrow to locate CRSFs. As a result, other ideas to separate the population have been proposed based on precession of a super-critical disc/wind (\citealt{Middleton_2018_Lense_Thirring, Middleton_2019_Accretion_plane, Khan2022}), via their X-ray spectra (\citealt{Pintore2017_pop,Koliopanos2017_pop, Walton2018_pop, Gurpide2021_pop}) or by associating ULXs undergoing large changes in X-ray flux with a propeller state. The latter approach assumes that a given NS-ULX is close to spin equilibrium (a reasonable assumption given the enormous spin-up rates, e.g. \citealt{Bachetti_2020}) such that the co-rotation radius (R$_{\rm co}$) moves inwards and approaches the magnetospheric radius (R$_{\rm m}$). When this occurs, accretion within R$_{\rm m}$, i.e. through the magnetosphere and onto the neutron star itself, may be completely or partially halted, with the abrupt change in luminosity permitting an estimate for the dipole magnetic field strength of the neutron star (e.g. \citealt{Cui1997, Tsygankov2016}). Attempts have been made to locate propeller states in ULXs by searching for large (over an order of magnitude) changes in X-ray brightness (e.g. \citealt{2018_Earnshaw_propeller_MNRAS.476.4272E, 2019_Song_hunt_for_pulx}). However, such drops in brightness might also result from precession of the disc and wind (\citealt{Pasham_M82_precessing_disc, Middleton_2015ULX_modelpaper}), or an increase in the scale-height of obscuring material/narrowing of the wind-cone (\citealt{Middleton_2015ULX_modelpaper, Gurpide_2021b}). Indeed, the continued spin-up of some PULXs during faint-states appears to confirm our line-of-sight to the central regions can indeed change (e.g. \citealt{2019_Vasilopoulos_MNRAS.488.5225V, Fuerst_P13_2021}).

In this paper, we provide a description of propeller-induced luminosity changes based on super-Eddington accretion theory and show that the dipole magnetic field strength of the neutron star can be constrained based on reasonable assumptions. We also provide a qualitative description of how the X-ray spectra may change under a propeller transition and tentatively identify such transitions for a handful of ULXs, constraining the dipole field strength in NGC 1313 X-1.

\section{propeller states in locally super-critical ULXs}

In the following, we will limit ourselves to the condition that the disc reaches the Eddington limit locally outside of R$_{\rm m}$, at the spherisation radius given by $R_{\rm sph} \approx \dot{m}_{0}R_{\rm in}$, where $R_{\rm in}$ is assumed to be the ISCO (whether reached by the disc or not, i.e. in the absence of truncating magnetic fields) and $\dot{m}_{0}$ is the accretion rate at very large radius in Eddington units; this is typically the mass transfer rate from the donor star unless some mass is lost at large radius due to thermal winds  (\citealt{Middleton2022}). This condition (i.e. $R_{\rm sph} > R_{\rm m}$) which seems to be met in PULXs (e.g. \citealt{King_Lasota_2020_PULX_iceberg_emerges}), requires the dipole field to not be extreme (B $>10^{13-14}$G) or that $\dot{m}_{0}$ is not too low. In such an accretion flow, the disc is assumed to follow standard theory (and knows nothing about the compact object at large radius, see \citealt{Shakura_1973}) and launches winds to remain locally Eddington limited at all radii down to R$_{\rm m}$. From R$_{\rm m}$ down to the neutron star surface (at R$_{\rm ns}$), material is assumed to free-fall before forming a shock and dissipating the gravitational energy. It is highly likely that some of the material within the magnetospheric flow will be launched in a wind, however, this has not yet been explored in detail. We assume that the presence of winds (now directly observed in several ULXs: \citealt{Middleton2014, Middleton2015_winds, Pinto_2016Natur.533...64P, Pinto2017_NGC55, Pinto2020_1313campaign, Walton2016_FeK_wind, 2018_Kosec_MNRAS.479.3978K, Kosec2021}) acts to collimate radiation generated within R$_{\rm sph}$. The collimation will be a function of radius (and $\dot{m}_{0}$, cf \citealt{King_2009_Beaming, Jiang_2014, Jiang_2019}), but here we will assume for simplicity that it can be parameterised at all radii within the wind-cone by b = $\Omega/4\pi$, where $\Omega$ is the solid opening angle of the wind (see \citealt{King_2009_Beaming}). We also note that the observational impact on collimation is naturally system inclination-dependent (e.g. \citealt{Dauser_2017_ULXLC,Middleton2021_SS433}).  

\subsection{Propeller states - without advection}

\begin{figure*}
    \centering
    \includegraphics[width=0.99\textwidth]{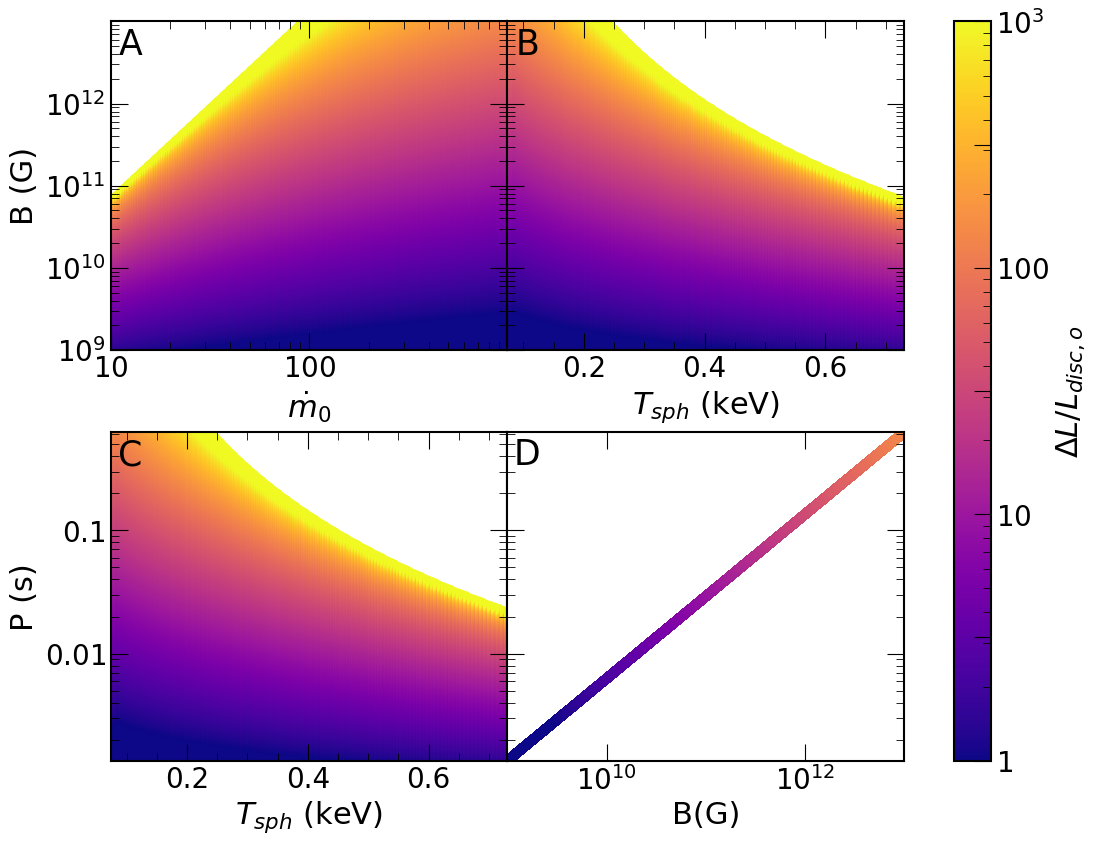}
    \caption{Plots showing the dependence of $\Delta L/L_{\rm disc, o}$ on the dipole magnetic field strength, B, and the accretion rate at large radius in Eddington units, $\dot{\rm m}_{0}$ (Panel A). The latter may alternatively be parameterised by the temperature at the spherisation radius, $T_{\rm sph}$ (Panel B), whilst the dipole magnetic field strength is related to the neutron star spin period at equilibrium (Panels C \& D). These calculations have assumed no explicit advection but a loss of energy to the wind, parameterised by $\epsilon_{\rm wind}$ = 0.25.}
\end{figure*}

We assume that binding energy liberated between $R_{\rm m}$ and the neutron star surface emerges either directly as pulsed emission, or via reprocessing in the magnetosphere (\citealt{Mushtukov_2017_opt_thick_env_NS}). The sum total luminosity of this radiative component is therefore limited to:

\begin{equation}
    L_{\rm ns} = GM_{\rm ns}\dot{M}_{m}\left[\frac{1}{R_{\rm ns}} - \frac{1}{R_{\rm m}}\right]
\end{equation}

\noindent where $R_{\rm ns}$ is assumed to be 10km, and $\dot{M}_{\rm m}$ is the mass accretion rate at $R_{\rm m}$ (and where we are explicitly assuming no mass loss between $R_{\rm m}$ and the neutron star surface even though this might well occur in practice). Assuming the opening angle of the radiating surface (both magnetosphere and star) subtends a larger solid angle than the wind opening angle, it will be collimated by the wind-cone to yield an observed luminosity of:

\begin{equation}
    L_{\rm ns, o} = \frac{GM_{\rm ns}\dot{M}_{\rm m}}{b}\left[\frac{1}{R_{\rm ns}} - \frac{1}{R_{\rm m}}\right]
\end{equation}

We expect an intrinsic accretion luminosity from the disc between R$_{\rm m}$ and R$_{\rm sph}$ of:

\begin{equation}
    L_{\rm disc} = L_{\rm Edd}\left[\ln\left({\frac{R_{\rm sph}}{R_{\rm m}}}\right) + 1\right]
\end{equation}

Where the ``+1" refers to emission from the relatively thin disc beyond $R_{\rm sph}$, which emits close to the Eddington limit (although flux from this region is likely to be intercepted and re-radiated from the photosphere of the wind: \citealt{Poutanen_2007_ln}). In practice regardless of radial advection the luminosity detected as radiation will be reduced by a factor $\xi$ = 1-$\epsilon_{\rm w}$ (where $\epsilon_{\rm w}$ is the fraction of the accretion luminosity used in driving the wind: \citealt{Poutanen_2007_ln}). The value of $\epsilon_{\rm w}$ is as-yet unknown (values vary between 0.25 from simulations: \citealt{Jiang_2014} to close to unity from observations of the densest phases of the wind: \citealt{Pinto_2016Natur.533...64P}). We assume that only the emission generated within $R_{\rm sph}$ is collimated, such that the observed luminosity from this region is then given by:

\begin{equation}
    L_{\rm disc, o} = L_{\rm Edd}\left[\frac{\xi}{b}\ln\left({\frac{R_{\rm sph}}{R_{m}}}\right) + 1\right]
\end{equation}

Before the propeller state is entered and accretion onto the neutron star is assumed to switch off, we expect to observe $L_{\rm ns, o} + L_{\rm disc, o}$, whilst in a propeller state we should see only the emission from outside of $R_{\rm m}$, i.e. $L_{\rm disc, o}$. This implies that, for accretion rates, where $\frac{\xi}{b}\ln\left({\frac{R_{\rm sph}}{R_{\rm m}}}\right) \gg 1$ and assuming that $b$ is similar for both components (supercritical disc and accretion column/magnetosphere), the beaming factor can be eliminated to leave:

\begin{equation}
    \frac{\Delta L}{L_{\rm disc, o}} = \frac{L_{\rm ns, o}}{L_{\rm disc, o}} \approx \frac{GM_{\rm NS}\dot{M}_{m}}{L_{\rm Edd}\xi\ln\left({\frac{R_{\rm sph}}{R_{m}}}\right)}\left[\frac{1}{R_{\rm NS}} - \frac{1}{R_{\rm m}}\right]
\end{equation}

\begin{figure*}
    \centering
        \includegraphics[width=0.99\linewidth]{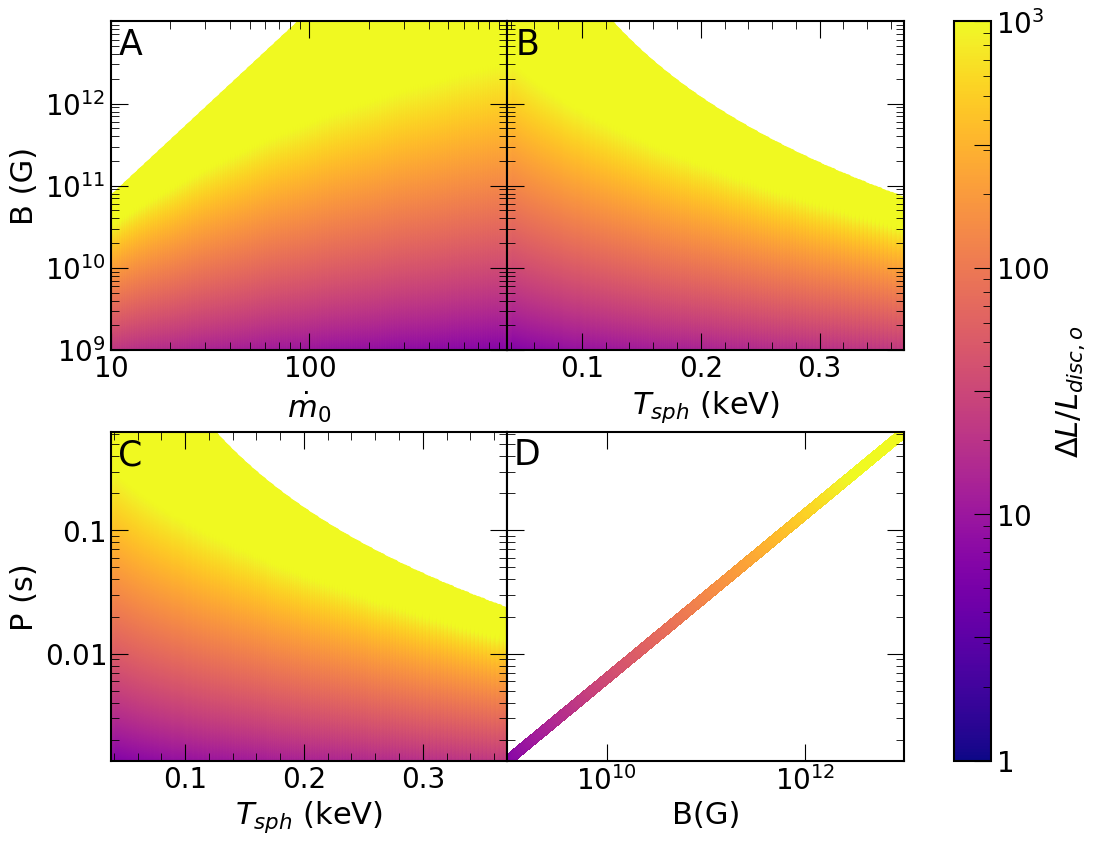} 
    \caption{As per Figure 1 but with  $\epsilon_{\rm wind}$ = 0.95.}
\end{figure*}

\noindent which, in the limit of high dipole field strength (such that $R_{\rm m} \gg R_{\rm ns}$), using canonical values and assuming the standard linear scaling of accretion rate with radius (i.e. neglecting advection: $\dot{m} = \dot{m}_{\rm 0}R/R_{\rm sph}$ \citealt{Shakura_1973}), can simplify to:

\begin{equation}
    \frac{\Delta L}{L_{\rm disc, o}} \approx \frac{0.208}{
    \eta} \frac{\dot{m}_{0}}{\xi} \frac{R_{\rm m}}{R_{\rm sph}}\left[\ln\left({\frac{R_{\rm sph}}{R_{m}}}\right)\right]^{-1}
\end{equation}

\noindent where $\eta$ is the radiative efficiency for accretion at the inner edge of the accretion disc (assumed to be $\sim 0.08$). In the above, $R_{\rm m}$ can be assumed to be related to the dipole field strength (B) via standard relations modified for super-Eddington accretion flows (i.e once again assuming a simple linear radial dependence for the accretion rate: \citealt{Shakura_1973}). Assuming a canonical neutron star mass of 1.4M$_{\odot}$) gives:

\begin{equation}
    R_{\rm m} \approx 4.2 \times 10^{7}B_{12}^{4/9}~~~~[{\rm cm}] 
\end{equation}

\noindent where $B_{12}$ is the dipole magnetic field strength in units of 10$^{12}$~G (and the magnetic dipole moment, $\mu = BR_{\rm ns}^{3}$). It is clear from equations 5--7, that, for $R_{\rm m} < R_{\rm sph}$, the change in luminosity during a transition to a propeller state, relative to the luminosity during a propeller state is a function of $\dot{m}_{0}$ and B (and an assumed value for $\xi$). Using equations 5 \& 7 (which allows us to cover the full range of dipole field strengths), we plot the resulting values for $\Delta L/L_{\rm disc, o}$ in Figures 1 and 2 for B between 10$^{9}$-10$^{13}$ G, $\dot{m}_{0}$ between 10 and 1000 and assuming values for $\epsilon_{\rm w}$ of 0.25 and 0.95 respectively. We note that we exclude those cases where $R_{\rm m}$ dips below the location of the ISCO (assumed to be 6 $R_{\rm g}$) within which the nature of the coupling is unclear.

Figure 1 (where $\epsilon_{\rm w}$ = 0.25) shows that, whilst substantial changes in luminosity are certainly possible for low to moderate accretion rates and high dipole field strengths, it is clear that a drop in total luminosity of less than a factor 10 can also occur for a wide range of dipole field strengths (between 10$^{9}$ and 10$^{11}$G coupled with a range in $\dot{m}_{0}$ between 10 and 1000). For those face-on and extremely bright sources, the source may well remain a very bright ULX even when in the propeller state. The same is true for Figure 2 (where $\epsilon_{\rm w}$ = 0.95) but to a lesser degree; as expected, less energy escapes as radiation from the disc and $\Delta L/L_{\rm disc, o}$ is therefore larger for a given $\dot{m}_{0}$ and B.

As the mass of the compact object in this picture is assumed to be known within a small range, $\dot{m}_{0}$ can be estimated from the equation of \citealt{Poutanen_2007_ln}: $T_{\rm sph} \approx 1.5 f_{\rm col}m^{-1/4}\dot{m}_{0}^{-1/2}~{\rm keV}$ where $f_{\rm col}$ is the colour temperature correction factor (for which we assume a value of 1.8). As a characteristic temperature can be easily obtained from soft X-ray spectra of ULXs (which will not be confused with emission from the magnetosphere when the field is only of moderate strength: \citealt{Mushtukov_2017_opt_thick_env_NS}, see Section 3), in principle, this allows the dipole field strength to be estimated directly from the combination of X-ray spectra and relative change in luminosity (although see the Discussion for issues related to obtaining an accurate value for $T_{\rm sph}$). The relative changes in the luminosity associated with the onset of a propeller state versus $T_{\rm sph}$ and B are also shown in Figures 1 \& 2.

Given the requirement that $R_{\rm m}$ is equal to the co-rotation radius at the point where the propeller transition occurs, we are also able to predict the spin period of the neutron star ($P_{\rm ns}$) through equating $R_{\rm m}$ to the co-rotation radius given by $R_{\rm co} = (GM_{\rm ns} P_{\rm ns}^{2}/4\pi^{2})^{1/3}$: 

\begin{equation}
P_{\rm ns} \approx 0.16 \frac{B_{\rm 12}^{2/3}}{\sqrt{m}}~~~[{\rm s}]  
\end{equation}

\noindent where $m$ is the neutron star mass in units of M$_{\odot}$. The equivalent plots of $B$ vs $P_{\rm ns}$ and $T_{\rm sph}$ vs $P_{\rm ns}$ are shown in Figures 1 \& 2 respectively. 

\begin{figure*}
    \centering
        \centering
        \includegraphics[width=0.99\linewidth]{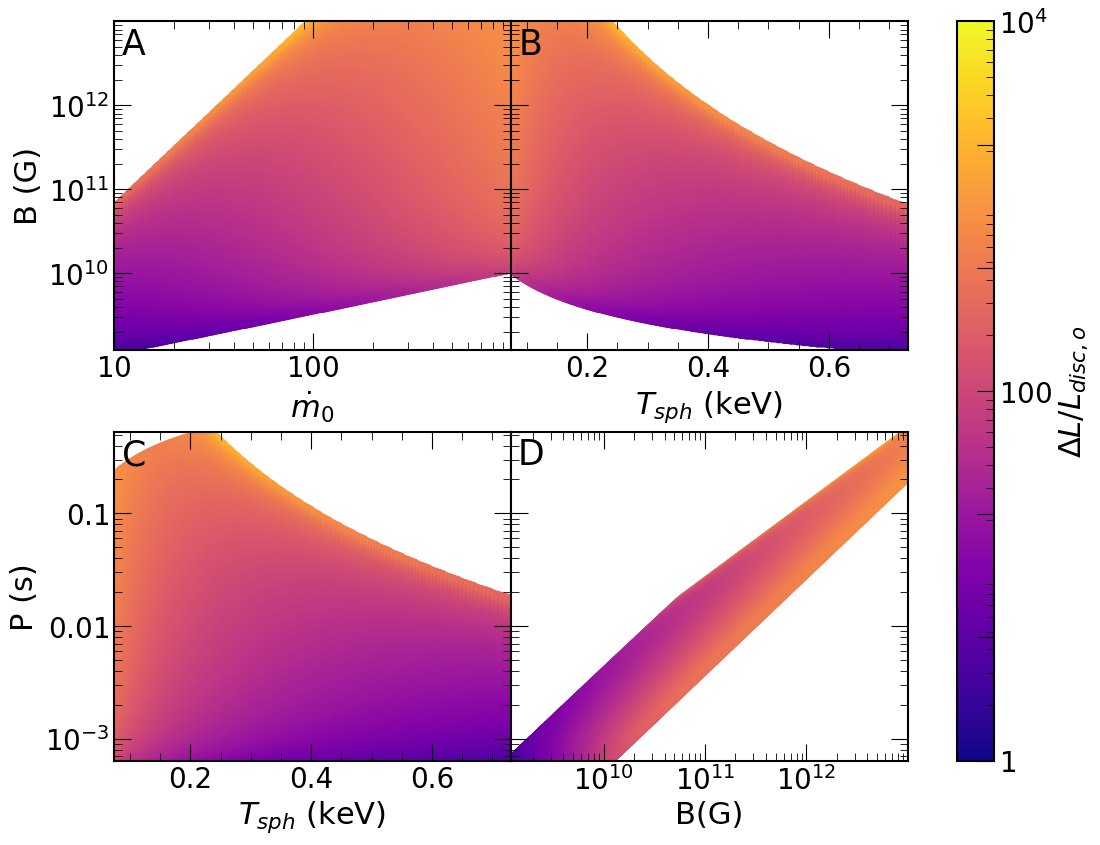}
 
    \caption{As per Figure 1 with explicit advection (see \citealt{Poutanen_2007_ln, Middleton_2019_Accretion_plane}) and a loss of energy to wind parameterised by $\epsilon_{\rm wind}$ = 0.25.}
\end{figure*}

\subsection{Propeller states - with advection}

The previous subsection assumed no explicit advection of matter and radiation within the disc, with only a fraction of the accretion power lost in driving in the wind (\citealt{Poutanen_2007_ln}). Whilst the true impact of advection on the cooling of the super-critical flow is still unclear (e.g. magnetic buoyancy may help radiation escape: \citealt{Jiang_2014}), as discussed in \cite{Chashkina2019}, including a radial advection term leads to a higher accretion rate at $R_{\rm m}$ which increases the accretion rate onto the neutron star at which point the energy must be released. 

Following \cite{Poutanen_2007_ln} and \cite{Middleton_2019_Accretion_plane}, we include the effect of advection by setting the constant $a = \epsilon_{\rm w}(0.83 - 0.25\epsilon_{\rm w})$ which enters the radial profile of $\dot{m}$ (noting that this differs to the simple profile assumed in the previous sub-section: \citealt{Poutanen_2007_ln}), which we then solve numerically to obtain $R_{\rm m}$ (see \citealt{Middleton_2019_Accretion_plane} for details). The equivalent plots when including advection are shown in Figs 3 \& 4 for values of $\epsilon_{\rm w}$ of 0.25 and 0.95 respectively (noting that the formula for $P_{\rm ns}$ also now depends on the advective term). As expected, the result of advection is a decrease in brightness of the accretion disc (as energy is advected radially inwards) and a corresponding increase in brightness of the regions within $R_{\rm m}$ (assuming the neutron star can actually accommodate such rates of infall, otherwise mass loss must ensue). Together, this results in an increase in $\Delta L/L_{\rm disc, o}$ compared to the case without explicit advection (cf Figures 1 \& 2).

\begin{figure*}
    \centering
        \includegraphics[width=0.99\linewidth]{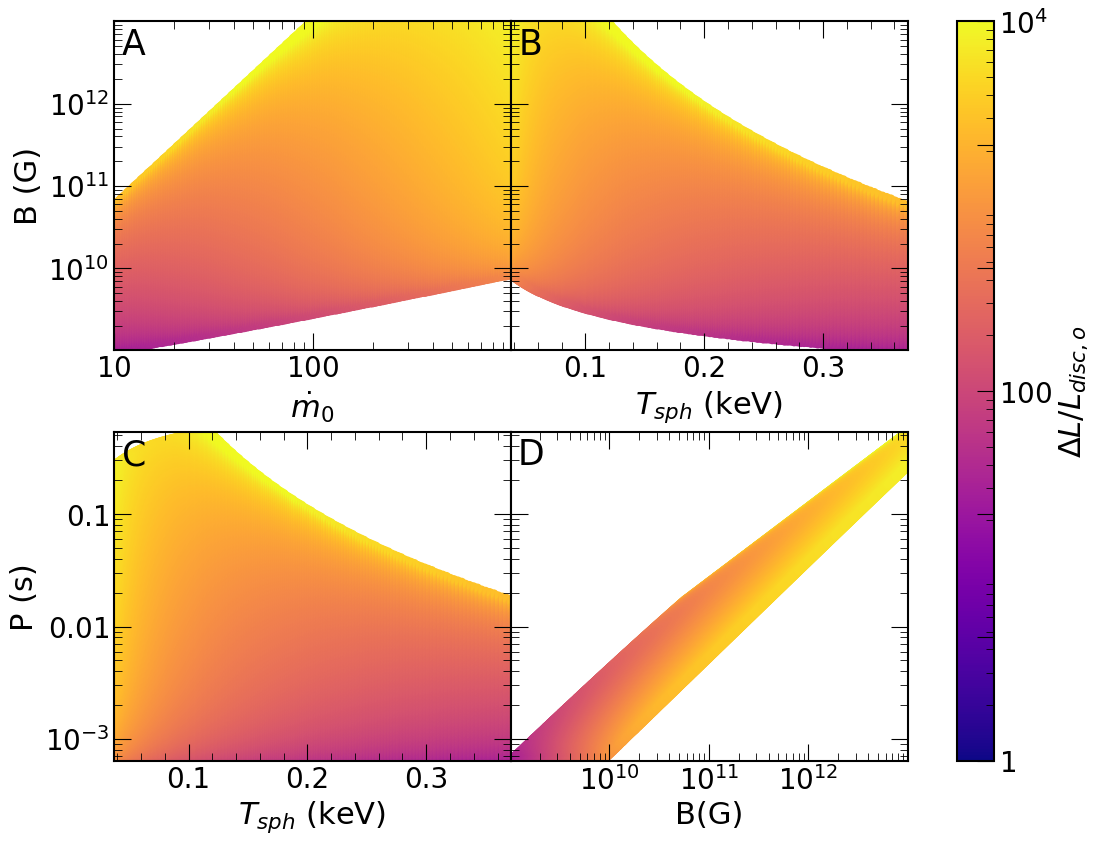} 
    \caption{As per Figure 3 but with $\epsilon_{\rm wind}$ = 0.95.}
\end{figure*}

\begin{figure*}
    \centering
        \includegraphics[width=0.49\linewidth]{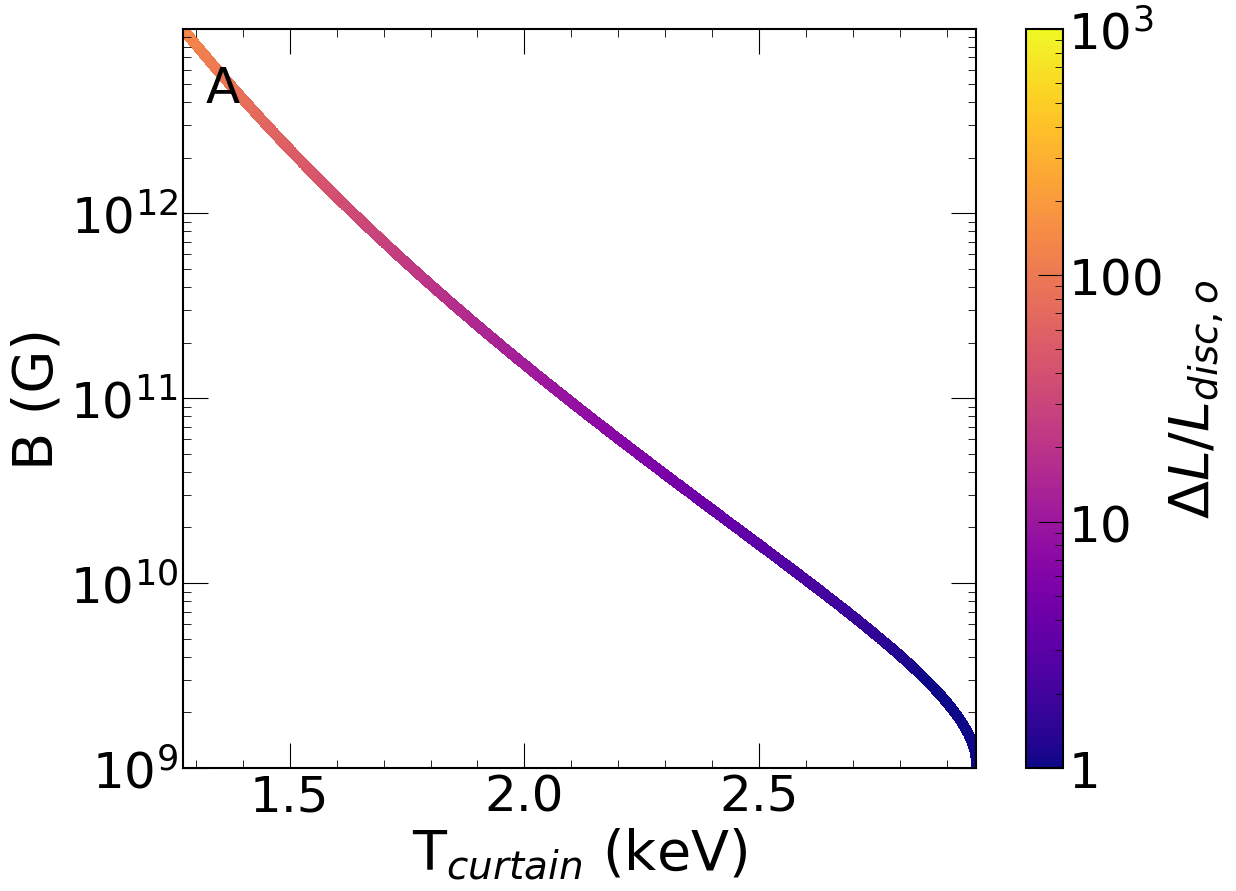} 
        \includegraphics[width=0.49\linewidth]{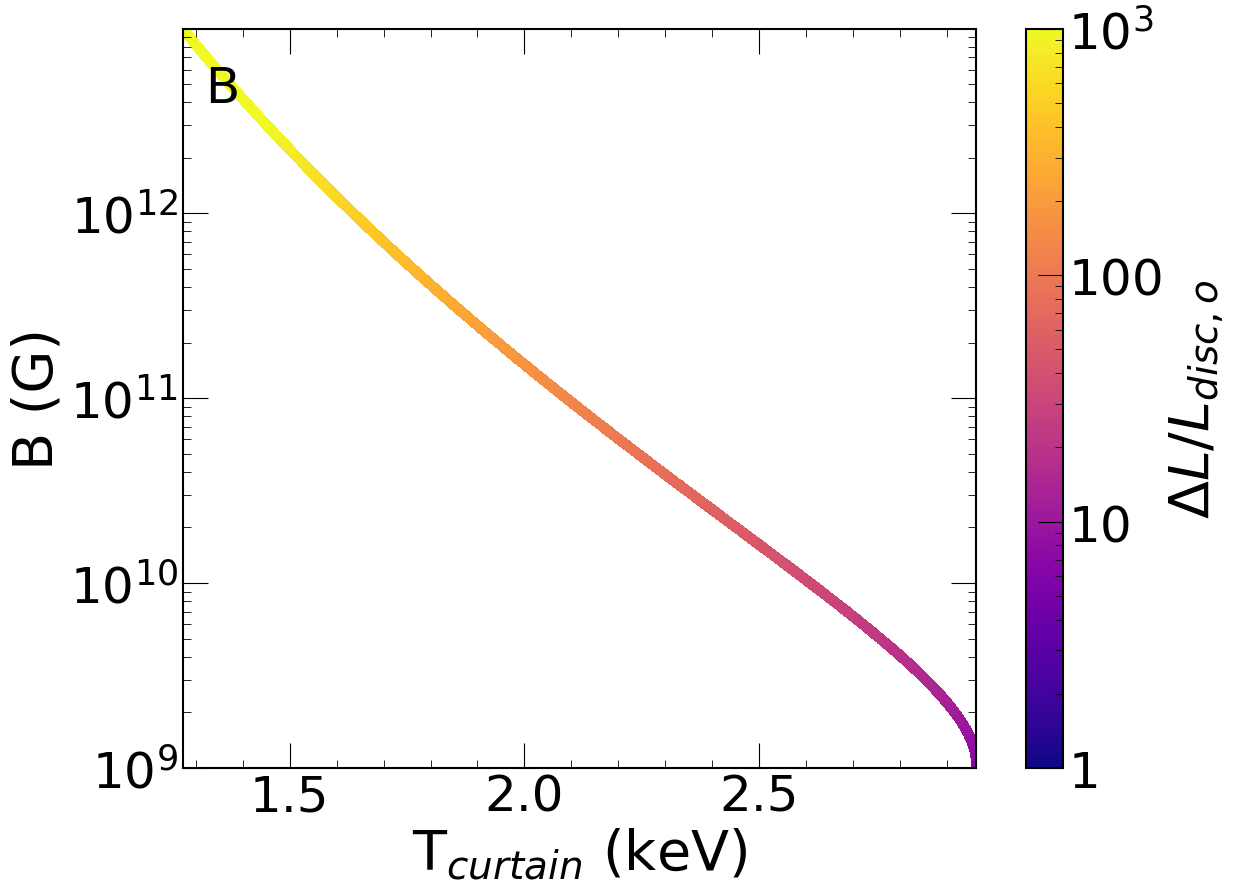} \includegraphics[width=0.49\linewidth]{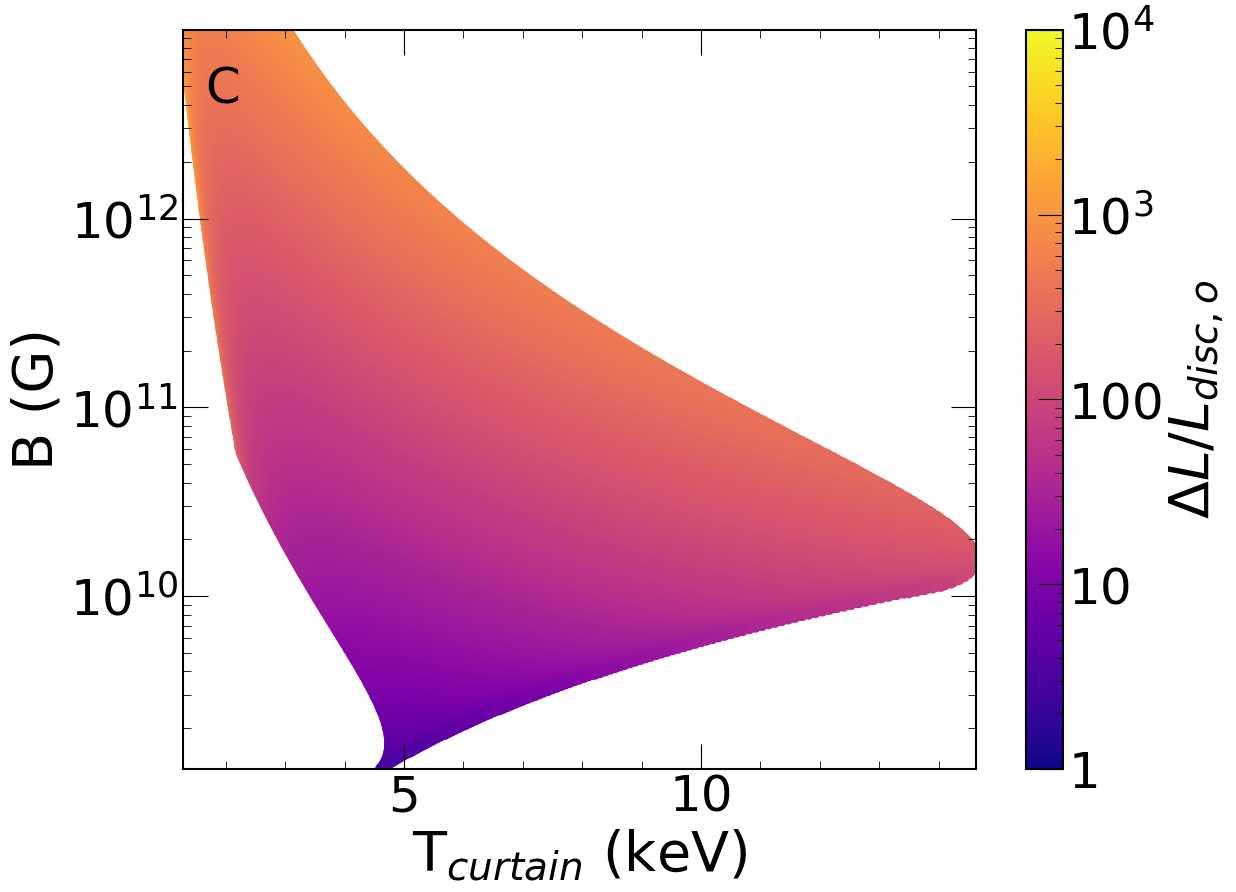}
        \includegraphics[width=0.49\linewidth]{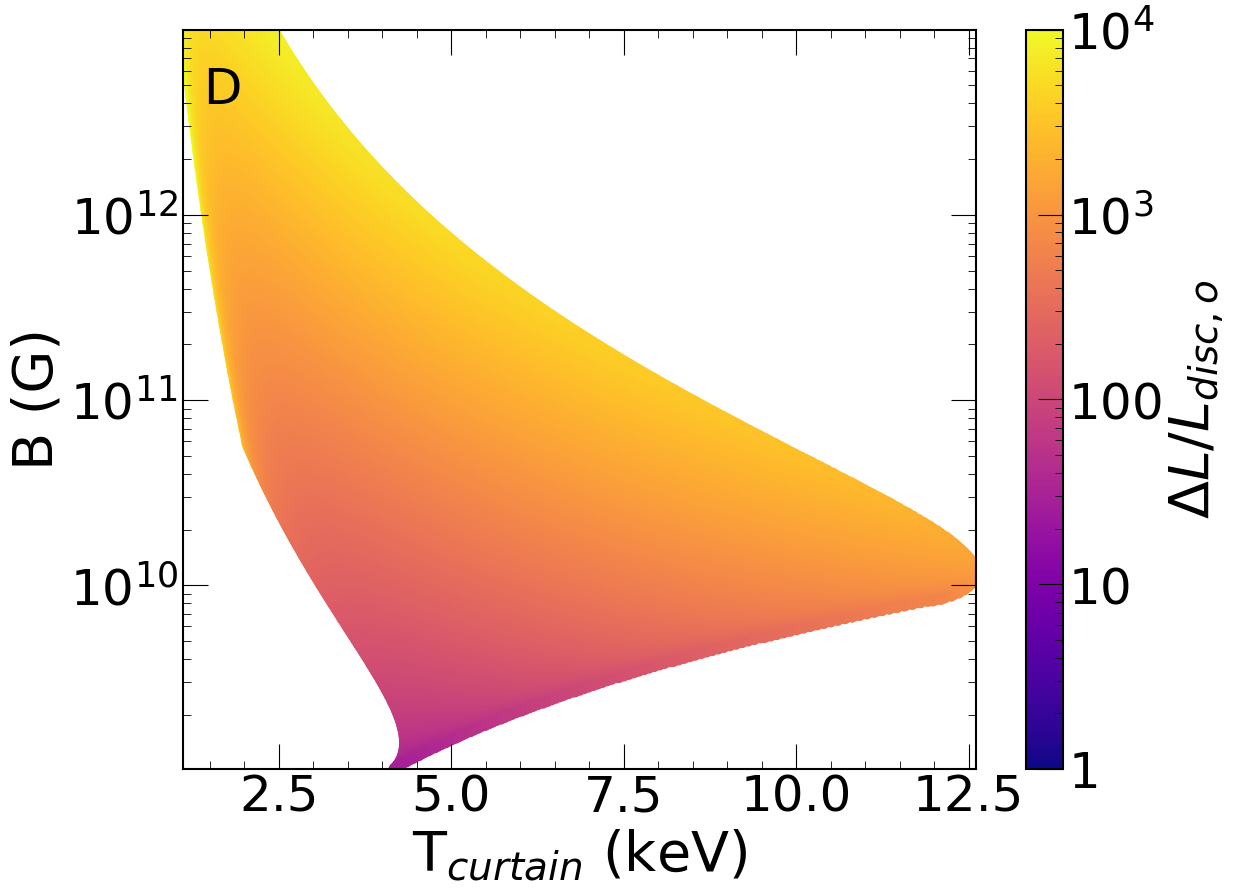} 
    \caption{Plots showing the dependence of $\Delta L/L_{\rm disc, o}$ on the dipole magnetic field strength, B, and the temperature of the accretion curtain (see \citealt{Mushtukov_2017_opt_thick_env_NS}). Panels A \& B assume no advection with $\epsilon_{\rm wind}$ = 0.25 and 0.95 respectively, whilst Panels C \& D have assumed explicit advection (see \citealt{Poutanen_2007_ln, Middleton_2019_Accretion_plane}) with  $\epsilon_{\rm wind}$ = 0.25 and 0.95 respectively.}
\end{figure*}

\section{Searching for possible propeller states in ULXs}

It is crucial to note that, in order to determine the dipole field strength using the above formulae, we require a bandpass where the accretion column (reprocessed or direct) can be seen to switch off. As noted by \cite{Brightman2016_M82} and \cite{Walton2018_P13}, the pulsed component can be directly and unambiguously identified in PULXs and dominates above a few keV. This component may be present in other ULXs, even where pulsations have not been observed (\citealt{Pintore2017_pop, Walton2018_pop}). A natural corollary is that a propeller state will be associated with this high energy component -- noted to remain stable across many observations in some ULXs (e.g. \citealt{Walton2017_HoIX, Walton_1313_2020, Gurpide2021_pop}) -- dropping away. In addition to the accretion column, we would also expect any emission from within R$_{\rm m}$, i.e. from the accretion curtain, to also fall away, however, importantly, the entire broad-band flux, especially at low energies does not need to fall to zero. To confirm the range of temperatures we would expect from the accretion curtain, we employ the formulae of \cite{Mushtukov_2017_opt_thick_env_NS} assuming the curtain is optically thick throughout and able to efficiently reprocess all of the emission from the fan-beam within (noting that complete reprocessing is a limiting assumption). The emergent temperature is altered by some colour temperature correction factor ($f_{\rm col}$ which we again set to a value of 1.8) and is given by the formula:

\begin{equation}
    T_{\rm cur} = kf_{\rm col}\left(\frac{L_{\rm ns}}{4\pi R_{\rm m}^{2}\sigma}\right)^{1/4}~~[keV]
\end{equation}

\noindent where $k$ is the Boltzmann constant, $\sigma$ is the Stefan Botltzmann constant, and $L_{\rm ns}$ is the intrisic luminosity generated from free-fall onto the neutron star (i.e. $bL_{\rm ns, o}$). For the ranges of dipole field strength and $\dot{m}_{0}$ we explore here, we plot the corresponding value of $T_{\rm cur}$ in Figure 5. It is apparent that the peak of the emission ($\sim$3 T$_{\rm cur}$) will be above 2-3 keV in all cases we are considering. This component is therefore unlikely to be readily confused with the soft component seen in most ULXs (e.g. \citealt{Middleton_2015ULX_modelpaper}) unless only a very small fraction of the fan beam luminosity is reprocessed (which would tend to conflict with the inferred luminosity of the soft component).

\begin{figure*}
    \centering
    \includegraphics[width=0.32\textwidth]{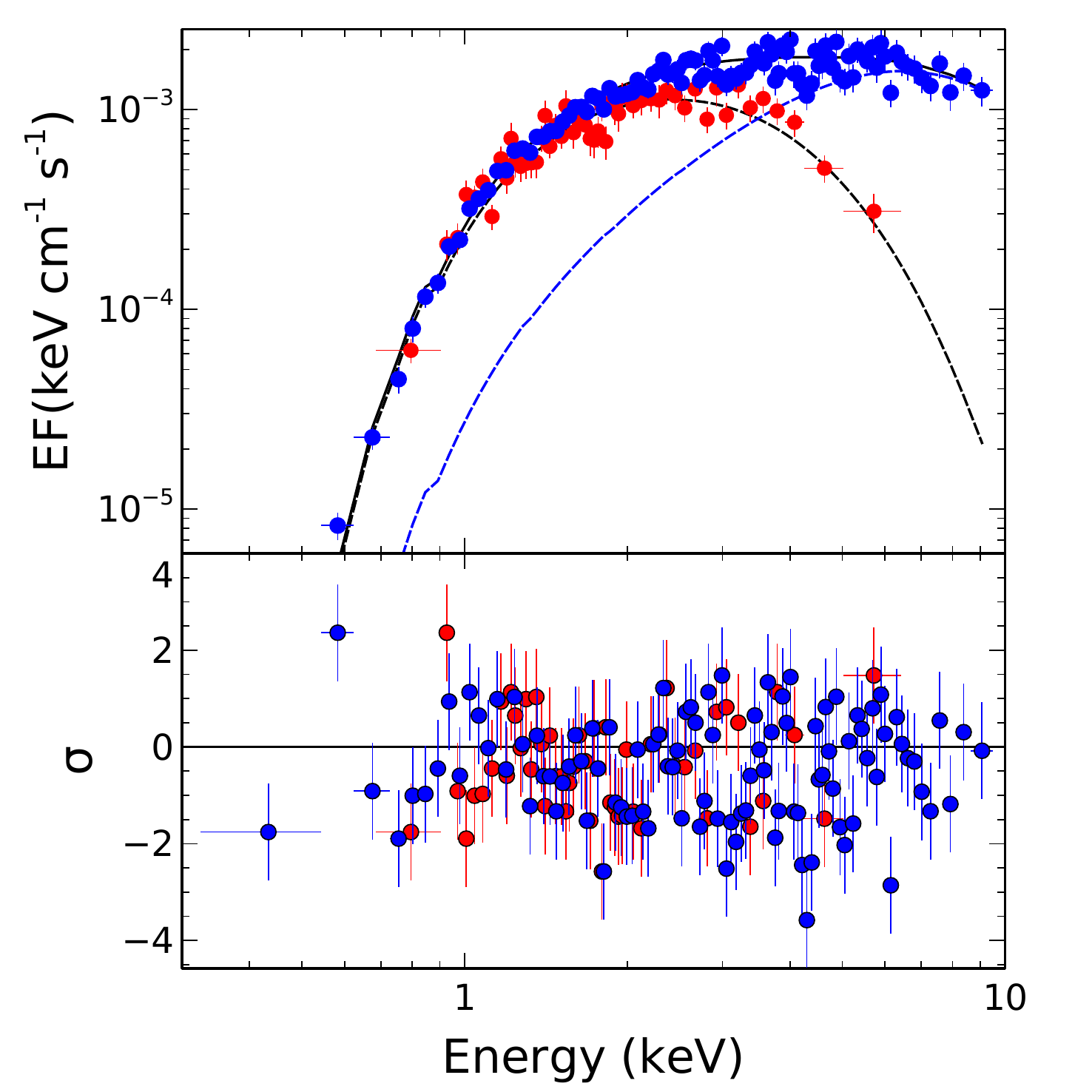}
        \includegraphics[width=0.32\textwidth]{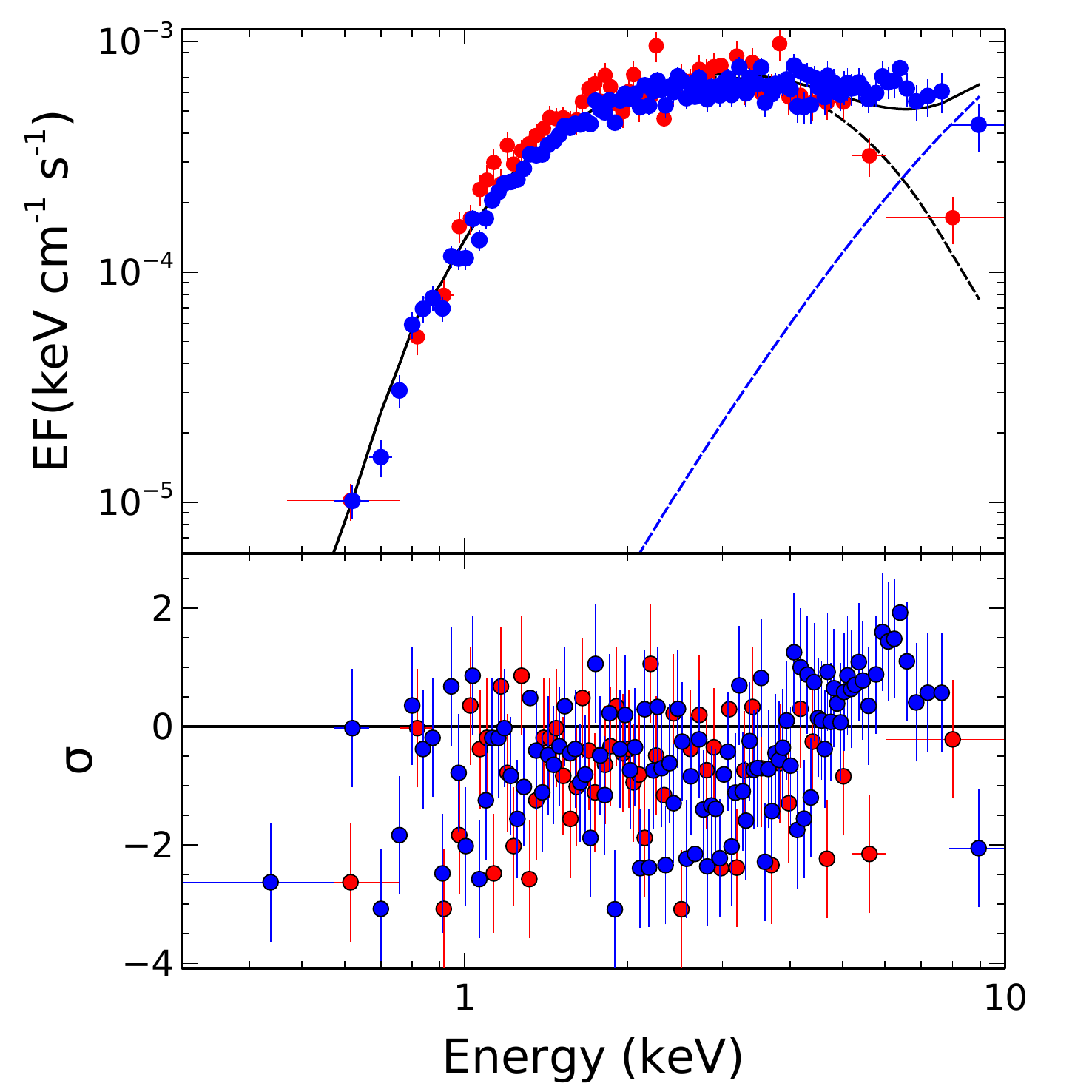}
            \includegraphics[width=0.32\textwidth]{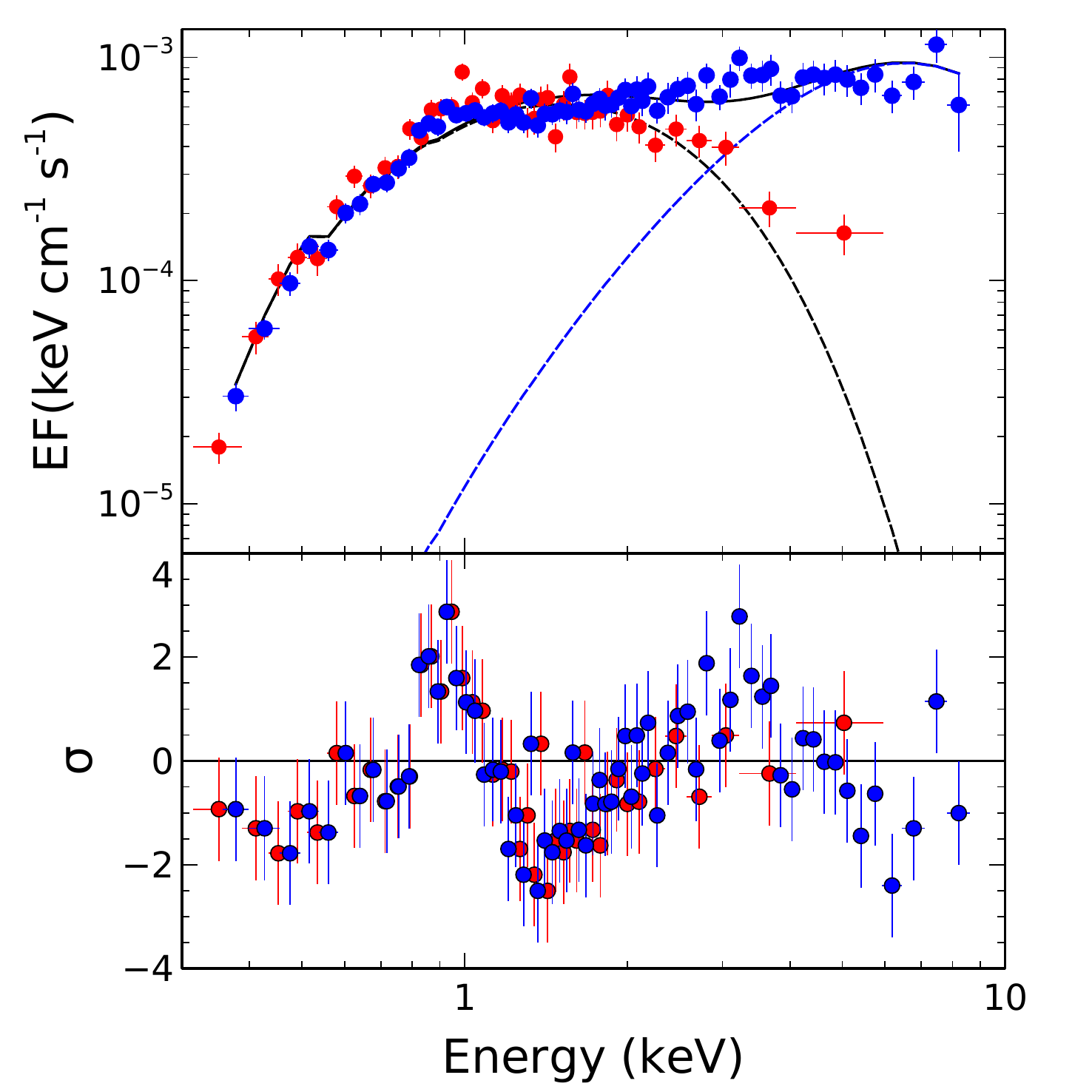}
    \caption{Spectral transitions tentatively attributed to propeller transitions. From left to right, the panels show IC 324 X--1, Cir ULX-5 and NGC 1313 X-1. All panels show \textit{XMM-Newton} EPIC PN data, except for the low state of IC 324 X--1 and the high state of Cir ULX-5, where \textit{Chandra}-ACIS and EPIC MOS1 data are shown instead. The high state and low (propeller) states are colored in blue and red respectively. The \texttt{diskbb} component (which is tied between states) is shown by a black dashed line, the \texttt{nthcomp} by a dashed blue line and the total model by a solid black line. Data has been rebinned for the purpose of clarity.}
    \label{fig:propellers}
   
\end{figure*}

Based on the above argument, ULXs where $R_{\rm m} < R_{\rm sph}$, entering the propeller regime should remain relatively bright in the soft band, while the emission above $\sim$ 2 keV should drop significantly. This scenario may be able to explain the unusual spectral states found in IC 342 X-1 (\citealt{Marlowe2014, Gurpide2021_pop}) where it appears that, between two observations separated in time, the unabsorbed luminosity in the soft band has remained stable but the hard band luminosity has dropped by a factor $\sim$ 2 (\citealt{Gurpide2021_pop}). Interestingly, this source has been proposed as a good NS-ULX candidate based on its hard X-ray spectrum (\citealt{Pintore2017_pop, Gurpide2021_pop}). \cite{Gurpide2021_pop} noted similar transitions in NGC~1313~X--1 (see also \citealt{Middleton_2015ULX_modelpaper}) and Circinus ULX 5, where the hard flux also dropped by a factor $\sim$2--3, with stable flux in the soft band. We revisit these spectral changes and, using the above formulae, attempt to provide initial estimates for the dipole magnetic field strength under the assumption that these spectral changes are caused by a transition to the propeller regime. More specifically, we used \textit{XMM-Newton} observations 0206890101, 0205230601, 0111240101 for the high state spectra of IC 342 X--1, NGC~1313~X--1 and Circinus ULX 5 respectively, and \textit{Chandra} observation 13686 and \textit{XMM-Newton} observations 0205230401, 0792382701 for the suspected propeller states of IC 342 X--1, NGC~1313~X--1 and Circinus ULX 5. 

The spectral products were taken from \cite{Gurpide2021_pop} and modelled in \texttt{XSPEC} version 12.12.1 using  \texttt{tbabs}$\otimes$(\texttt{diskbb} + \texttt{nthcomp}) following \cite{Middleton_2015ULX_modelpaper} (and using abundances and cross sections from \citealt{Wilms2000_tbabs} and \citealt{Verner1993, Verner1995}). Given the above picture, we require the soft component to be fairly stable between observations, whereas the hard component should have dropped in the propeller state. Therefore, we simultaneously fitted the two sets of spectra for each ULX, tying the luminosity of the \texttt{diskbb} and removing the \texttt{nthcomp} (the emission within Rm) from the propeller state. The absorption column is also tied between observations, 
with the lower limit set at the respective line-of-sight column densities. We computed de-absorbed, model fluxes between 0.01 and 50 keV (to capture the total flux) using the pseudo-model {\sc cflux} for each of the model components (the \texttt{diskbb} and the \texttt{nthcomp}). It is important to note that the thermal tail of the \texttt{nthcomp} component may lead to a slight underestimate of the total flux above 10 keV (where the true emission is flatter than Wien, see e.g. \citealt{Walton_1313_2020}). For the sake of brevity, we present only the most relevant spectral parameters in Table 1, and direct the reader to the literature for a more complete picture of the spectral analysis of these sources (e.g. \citealt{Marlowe2014, Middleton_2015ULX_modelpaper, Gurpide2021_pop}). The spectral transitions between the high and putative propeller states and the best-fitting models are shown in Figure~\ref{fig:propellers}.

\begin{table*}
    \caption{Best-fit parameters for the putative propeller transitions. The model used for the high state was \texttt{tbabs}$\otimes$(\texttt{diskbb} + \texttt{nthcomp}), while for the propeller state the same model was used, but without the \texttt{nthcomp}. $L_\text{disc, o}$ and $L_\text{ns, o}$ refer to the unabsorbed 0.01--50 keV luminosities of the  \texttt{diskbb} and \texttt{nthcomp} model components respectively. Uncertainties are at the 1$\sigma$ confidence level. Distances adopted for the luminosity calculations are: 3.39, 4.25 and 4.21 Mpc for IC 342 X--1, NGC 1313 X--1 and Circinus ULX5 respectively \citep{Tully_cosmicflows-3_2016}. }
    \centering
\begin{tabular}{lccccccc} 
 \hline \hline 
 \noalign{\smallskip} 
Source & State & $N_{\rm H}$ &  $kT_{\rm soft}$ & $L_\text{\rm disc, o}$ & $L_\text{\rm NS, o}$ & $\Delta L / L_\text{\rm disc, o}$ & $\chi^2$/dof  \\ 
 &  &  $\times$10$^{20}$ cm$^{-2}$  & keV &  $\times$10$^{39}$ erg/s & $\times$10$^{39}$ erg/s & \\ 
\hline
\noalign{\smallskip}
\multirow{2}{*}{IC 342 X-1} & High &  \multirow{2}{*}{0.80$\pm$0.02} &  \multirow{2}{*}{0.87$\pm$0.02} &   \multirow{2}{*}{5.9$^{+0.2}_{-0.1}$} &  5.1$\pm$0.2 &  \multirow{2}{*}{0.86$\pm0.04$} &  \multirow{2}{*}{476/417} \\ 
 & Propeller &   &   &    & -  &  &   \\ 
 \multirow{2}{*}{NGC 1313 X-1} & High &  \multirow{2}{*}{0.15$\pm$0.01} &  \multirow{2}{*}{0.57$\pm$0.01}  &  \multirow{2}{*}{4.7$^{+0.09}_{-0.07}$} &  4.2$\pm$0.2 &  \multirow{2}{*}{0.90$^{+0.05}_{-0.04}$} & \multirow{2}{*}{717/390} \\ 
 & Propeller &   &  &   & - &  &    \\ 
 \multirow{2}{*}{Circinus ULX5} & High &  \multirow{2}{*}{0.63$\pm$0.01} &   \multirow{2}{*}{1.13$^{+0.01}_{-0.02}$} &   \multirow{2}{*}{5.15$\pm$0.05} &  8$^{+8}_{-3}$ &  \multirow{2}{*}{1.5$^{+1.6}_{-0.6}$} & \multirow{2}{*}{518/374}  \\ 
 & Propeller &   &  &  &  -  &  &  \\
\noalign{\smallskip} 
 \hline \hline 
 \end{tabular}

    \label{tab:my_label}

\end{table*}

Using the values in Table 1 and the formulae in this paper, we can attempt to place some constraints on the strength of the dipole field in the three ULXs assuming they contain neutron stars. Notably, all luminosities during the putative propeller state are above the Eddington luminosity for a neutron star, requiring $R_{\rm sph} > R_{\rm m}$ which matches the regime where our formulae should be reasonable approximations.

From Figures 1-4, it is apparent that the dominant uncertainty on any estimate for the dipole field strength via a propeller state is the role of advection and the value of $\epsilon_{\rm w}$. For a given $\Delta L/L_{\rm disc, o}$, the case with low $\epsilon_{\rm w}$ and no advection appears to place the most generous upper limit on the dipole field strength given a measurement of $T_{\rm sph}$. However, it is important to note that $T_{\rm sph}$ is currently estimated from a highly simplified model ({\sc diskbb}) which will be a poor description of the anisotropic super-critical disc and only return the highest temperature, not necessarily that from $R_{\rm sph}$ (instead, the temperature could correspond to emission from smaller radii); addressing this issue is beyond the scope of this paper but work is in progress to provide the community with more physically appropriate models. In the cases of IC 342 X-1 and Cir ULX-5, the implied column density is very high which may be further complicating our modelling of the soft emission. We therefore focus only on NGC 1313 X-1; given the values in Table 1 and the limits imposed by the case of no advection and low $\epsilon_{\rm w}$, we would estimate an upper limit of B $<$ 10$^{10}$ G unless the temperature estimated for T$_{\rm sph}$ is incorrect by over an order of magnitude.

\section{Discussion \& conclusions}

Searching for propeller states in ULXs is a promising means by which to locate neutron star accretors (e.g. \citealt{Earnshaw_ULX_cat, 2019_Song_hunt_for_pulx}). In this paper we have utilised the formulae for super-critical discs with mass loss and advection to explore the signatures of propeller states. In the case where radial advection is weak (such that only the local accretion luminosity escapes from within $R_{\rm m}$ and a simple linear scaling of accretion rate with radius is assumed: \citealt{Shakura_1973}), and a low fraction of the accretion power is used to drive the wind, propeller states with predicted drops in total luminosity of less than an order of magnitude exist across a substantial area of parameter space (in $\dot{m}_{0}$ and dipole field strength). Notably, even those large changes which accompany large dipole field strengths may still appear as (soft) ULXs when in a propeller state. Including radial advection (where the mass accretion rate now scales in a more complex manner: \citealt{Poutanen_2007_ln}) results in far larger changes in luminosity for a given dipole field strength, as relatively more accretion energy escapes at the neutron star surface (although it is likely that excess energy will be used to drive material from the magnetosphere). By combining the results with and without advection, it is clear that locating a source unambiguously entering a propeller state should allow constraints to be placed on the strength of the dipole magnetic field.

\cite{Tsygankov_M82_propellor} explored the dipole field strength implied by the factor $\sim$ 40 change in luminosity of M82 X-2 under the assumption of a propeller state where the disc beyond $R_{\rm m}$ is not locally super-critical, finding the field to be $\sim$10$^{14}$~G. Using our formulae, we would only obtain solutions where the spin period was faster than observed, implying that our assumptions have broken down (i.e. that $R_{\rm m} > R_{\rm sph}$) or that the change is not consistent with a propeller transition. Indeed, we now believe the changes in M82 X-2's flux to be driven by an entirely different effect (due to their periodic nature seeming to match other ULX superorbital periods  \citealt{Brightman2019M82period}). 

A key signature of a propeller state in those ULXs where $R_{\rm m} < R_{\rm sph}$, is a drop at hard X-ray energies above 2 keV (see Figure 5) as the column and curtain (\citealt{Mushtukov_2017_opt_thick_env_NS}) switch off, whilst the soft X-ray emission should remain essentially unchanged across the transition. Similar spectral changes might result from precession of the disc and wind (e.g. \citealt{Middleton_2015ULX_modelpaper, Middleton_2019_Accretion_plane}), however, in those cases we would expect some changes at soft energies as the beaming pattern has a radial dependence. This implies that searches which incorporate observations at high energies and hardness/flux ratios may be more efficient than simple searches for changes in luminosity (and will be better suited to locate the more subtle changes where the dipole field is weak). This line of reasoning provides additional evidence against major propeller-driven changes in M82 X-2, as the spectrum remains hard in the low luminosity state (\citealt{Brightman_2019_M82_period}). However, in the case of M82 X-2, true spin-down has already been observed (\citealt{Bachetti_2020}) without a major change in luminosity from the ULX. Unless the dipole field is weak, this implies material is still being fed at a substantial rate onto the neutron star and points to some natural complexity in how the field couples to the surrounding material. In such cases where the spectrum is unchanged -- and therefore not the ideal picture described above where accretion is completely halted within $R_{\rm m}$ -- the change in luminosity instead places a lower limit on the dipole field strength.

\begin{figure*}
    \centering
        \includegraphics[width=0.48\linewidth]{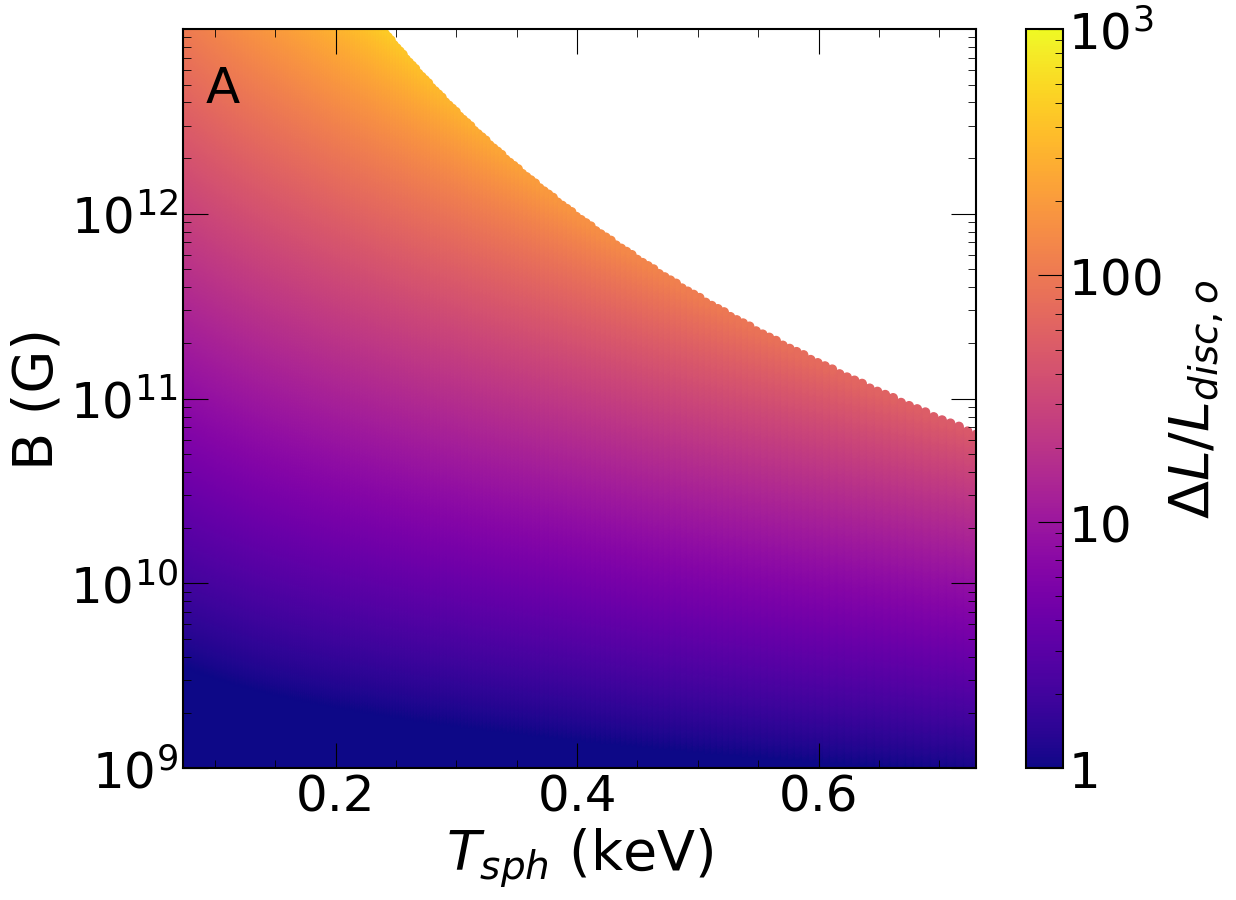} 
        \includegraphics[width=0.48\linewidth]{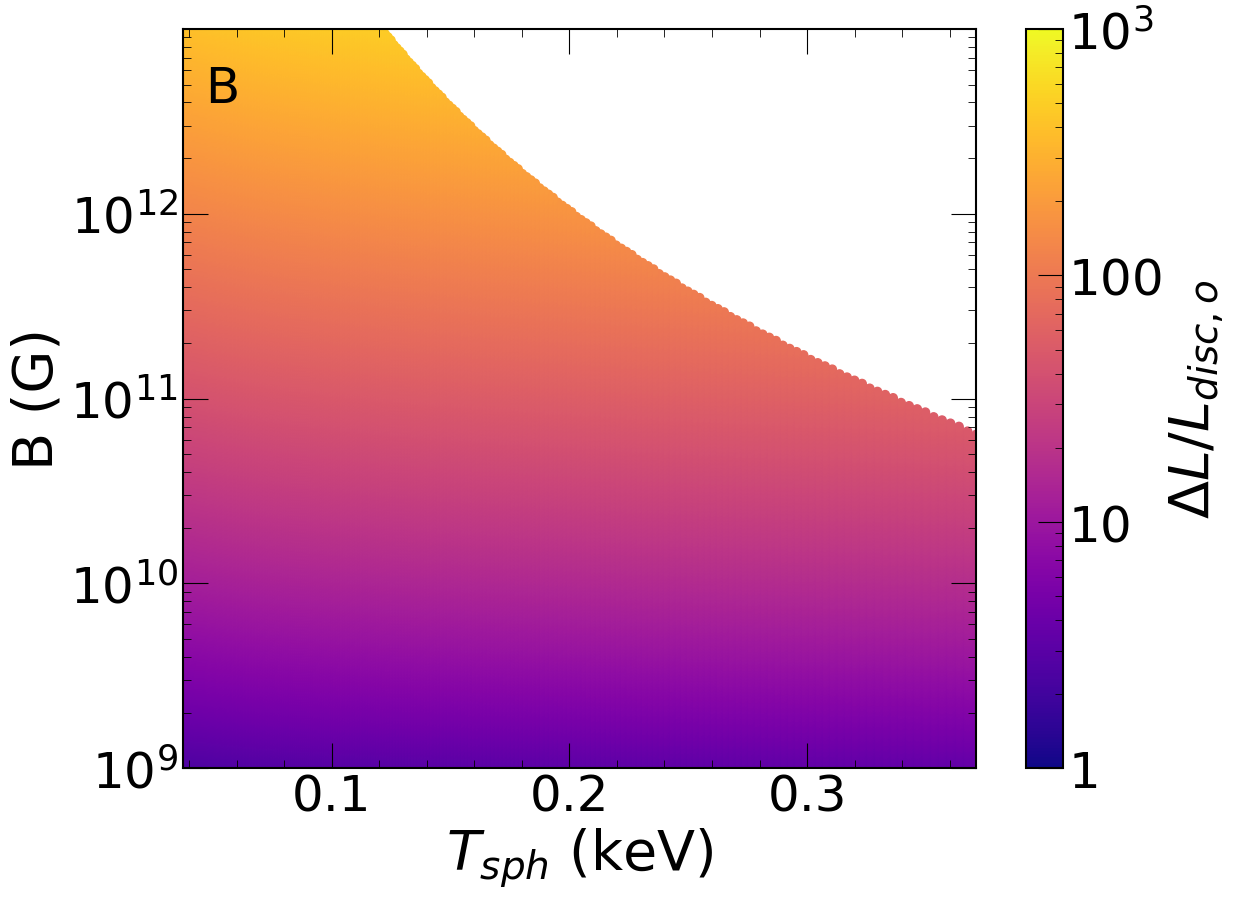} 
        \includegraphics[width=0.48\linewidth]{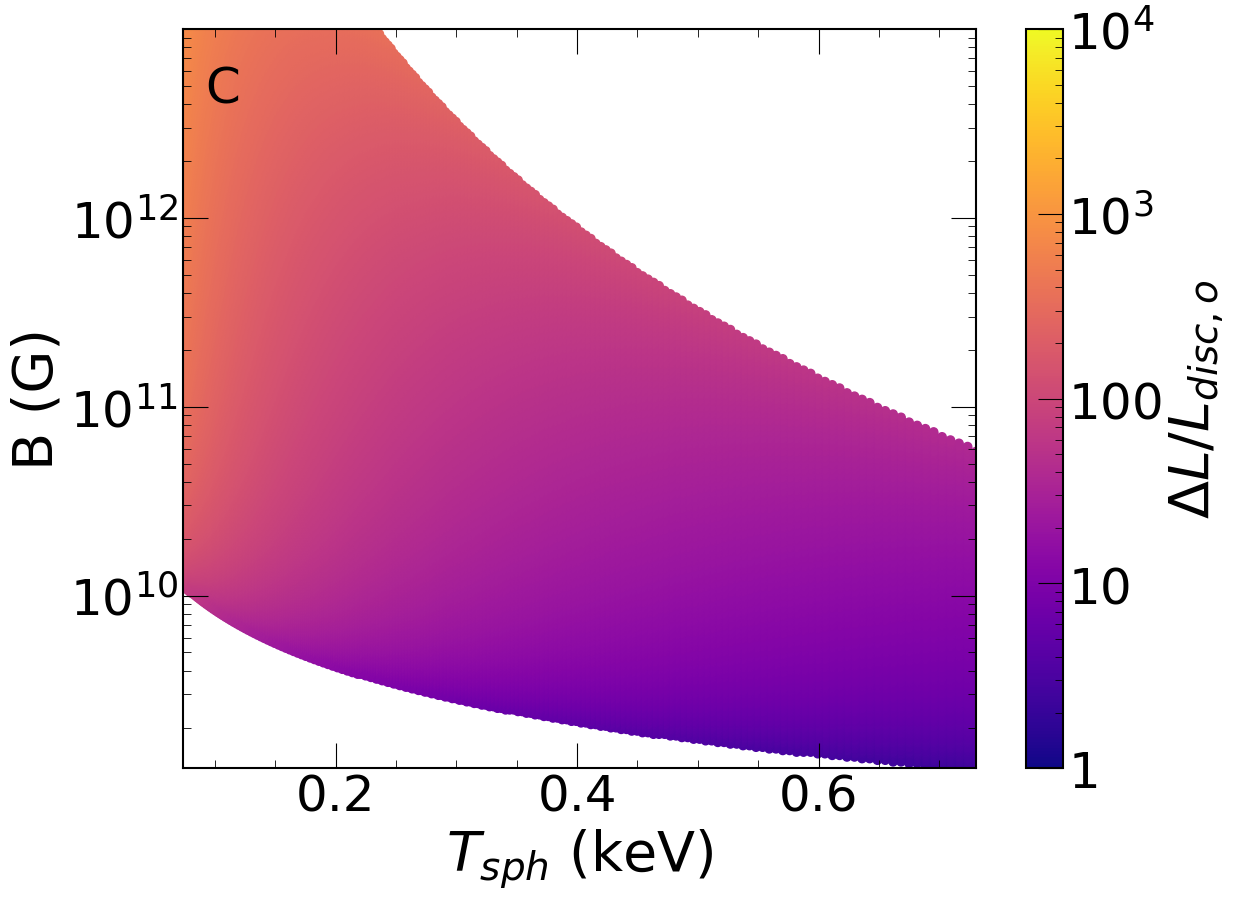} 
        \includegraphics[width=0.48\linewidth]{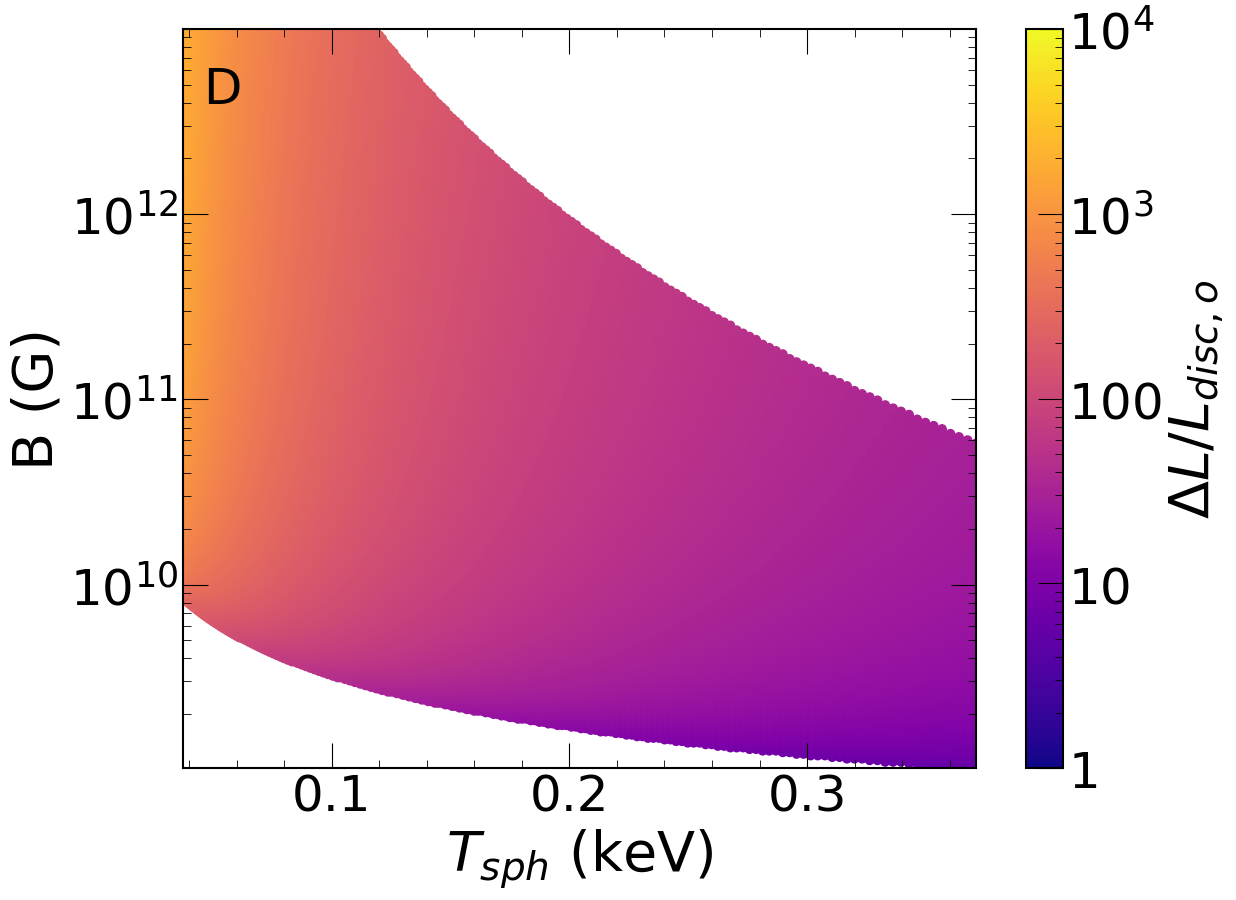}
    \caption{Plots showing the exact solutions for $\Delta L/L_{\rm disc, o}$ on $\dot{m}_{0}$, parameterised here by $T_{\rm sph}$ assuming $b$ = 0.5. Panels A \& B should be compared to Panels B of Figures 1 \& 2 whilst Panels C \& D should be compared to Panels B of Figures 3 \& 4.}
\end{figure*}

We have examined three ULXs where there appears to be spectral evidence in favour of the systems having entered a propeller state, IC 342 X-1, NGC 1313 X-1 and Circ ULX-5. All three ULXs show spectra where the hard component has fallen away (noting that this component in ULXs tends to have a similar shape at high energies: e.g. \citealt{Walton2017_HoIX}) leaving only a soft component. We are able to match the soft component to other observations of the same ULX and obtain some estimate for $\Delta L/L_{\rm disc, o}$. It is important to note that these observations are not contiguous (i.e.  we do not catch a transition into propeller) and so it is quite possible that there could be small changes in inclination and/or $\dot{m}_{0}$. Interestingly, in the case of NGC 1313 X-1, the CCD residuals -- now known to indicate the presence of a mass-loaded wind (\citealt{Middleton2014, Middleton2015_winds, Walton2016_FeK_wind, Pinto_2016Natur.533...64P, Kosec2021}) -- remain at a similar strength in both the high and propeller states, implying that winds of a similar strength are being driven even when emission from smaller radii is absent. The lack of change in CCD residual strength may well be an argument against precession producing this specific spectral change, as the strength of the atomic lines in NGC 1313 X-1 vary with spectral hardness as would be expected to accompany precession (\citealt{Middleton2015_winds, Pinto_1313_2020}).

The key parameters we can extract from the X-ray spectrum are a characteristic temperature at soft energies -- typically associated with the spherisation radius -- and the component luminosities in both states. In the cases of IC 342 X-1 and Circinus ULX-5, the soft emission is harder to model due to the large absorption columns and we obtain higher temperatures than the vast majority of ULXs (\citealt{Middleton_2015ULX_modelpaper}). Although the spectrum is easier to reliably model in NGC 1313 X-1, we note that estimates of T$_{\rm sph}$ from applying available models is highly uncertain and the mismatch between phenomenological models and the true anisotropy of super-critical discs is likely to produce overestimates for T$_{\rm sph}$. Regardless, we infer that in the most constraining condition of no advection and limited loss of accretion power to the wind, the dipole field in NGC 1313 X-1 -- should it harbour a neutron star -- should be $<$ 10$^{10}$G. Should this be correct, the corresponding neutron star rotation period should be very short ($\sim$ ms) and would be heavily diluted due to the signal being scattered in the wind cone (\citealt{2021_Mushtukov_MNRAS.501.2424M}). Such a weak dipole field strength is at odds with those measured for HMXBs in our own Galaxy (e.g. \citealt{Fuerst2014, Bellm_2014}) and even those other ULXs where the field is estimated from the spin-up rate (e.g. \citealt{NSULX_Furst_2016, King_2017_Pulsating_ULXs, Carpano_2018_NSULX_NGC300}), but {\it is} consistent with the recent study of a ULX in M 101 (4XMM J140314.2 + 541806) by \cite{Urquhart2022} where the dipole field was estimated to be of a similar strength ($\sim$ 10$^{10}$G) based on a possible propellor state transition. Such relatively low dipole fields could be consistent with the field being buried due to the high rates of accretion (e.g. \citealt{Geppert1994}) and may well be accompanied by stronger, higher order fields (\citealt{Israel1702, Israel1703, Middleton_Brightman_2019_M51, Brice2021, Kong2022_multipole}).

Should we be observing true propeller states in the ULXs studied in this paper, we must consider the process driving $R_{\rm co}$ to approach $R_{\rm m}$. Should the accretion rate be constant over time (noting that there are means by which to modulate this at large radius: \citealt{Middleton2022}), then the role of advection should be unchanged and $R_{\rm co}$ can only move inwards as the neutron star extracts angular momentum from the disc and spins up (noting that the accretion torque asymptotes to zero as $R_{\rm m}$ approaches $R_{\rm co}$ -- \citealt{Dai2006} -- demanding that some other process triggers the onset of the propellor state itself). This would imply that $R_{\rm co} \approx R_{\rm m}$ and we should not observe propeller transitions with any preference for spectral appearance, instead spectral changes would need to be driven by precession (\citealt{Middleton_2015ULX_modelpaper}). Alternatively, should $\dot{m}_{0}$ be changing with time, the impact of advection should also change (e.g. \citealt{Chashkina2019}). In the model we have applied, a larger $\dot{m}_{0}$, corresponds to a larger advected mass fraction and a smaller $R_{\rm m}$ for a given dipole field strength. For a fixed inclination angle where we can see into the wind-cone, a larger $\dot{m}_{0}$ would imply the source should be brighter due to increased collimation, whilst a smaller $\dot{m}_{0}$ -- where $R_{\rm m}$ is larger and can approach $R_{\rm co}$ -- should be fainter. Thus, should propeller states be driven mostly by changes in $\dot{m}_{0}$ rather than spin-up, we would naturally expect to see ULXs lose their hard component when they are in their faintest states. This is not what is observed in the ULXs we have explored in this paper where they appear bright (although not as bright as their peak observed luminosity, cf \citealt{Gurpide2021_pop}). This picture becomes more complicated when precession is involved, as the brightness and spectral shape are also dependent on phase (\citealt{Abolmasov2009, Middleton_2015ULX_modelpaper, Narayan2017_ULX}), implying that propeller transitions could occur without a preferred spectral shape and brightness; exploring this requires monitoring at soft to hard energies. 

An alternative possibility to explain changes in the spectra and brightness of the ULXs in this paper, is that we are not observing a propeller state but instead long duration dips, as material -- likely entrained in a wind -- passes across our line-of-sight (e.g. \citealt{Stobbart2004_NGC55, Dai2021_247, Alston2021_dips, Gurpide_2021b}, which then obscures the hard component of the spectrum. Although we do not detect obvious dips within the lightcurves of the putative propeller observations we have analysed, further distinguishing between these scenarios is beyond the scope of this current work.

There are a number of caveats to this work. In the absence of a radial beaming profile, we have assumed that collimation (determining the value of $b$ in equations 2 \& 4) affects the emission equally between $R_{\rm sph}$ and $R_{\rm m}$ and from within $R_{\rm m}$. In reality this is unlikely to be strictly true although becomes increasingly the case as the dipole field strength weakens or as the accretion rate increases. We also do not consider the probable mass loss between $R_{\rm m}$ and $R_{\rm ns}$ driven by the intrinsically super-Eddington flux from the column or as a result of the mass rate being fed onto the neutron star at larger rates than can be accommodated by the magnetic field (although this may be multipolar in nature and considerably stronger than the dipole: \citealt{Israel_2017_NSULX_5907, Tsygankov2017_multipole, Middleton_Brightman_2019_M51, Brice2021, Kong2022_multipole}). Should mass be lost from the magnetosphere, then the ratio of $\Delta L/L_{\rm disc, o}$ will be smaller for a given dipole field strength, allowing higher dipole field solutions; this remains an open question. In order to remove the beaming factor from our equations, we assumed  $\frac{\xi}{b}\ln\left({\frac{R_{\rm sph}}{R_{\rm m}}}\right) \gg 1$. This is only strictly correct for small $\epsilon_{\rm w}$, significant beaming (small $b$) or $R_{\rm sph}$ several factors larger than $R_{\rm m}$. Where this breaks down (when the beaming is not strong, which will occur as $R_{\rm m}$ tends to $R_{\rm sph}$), our estimate for the relative change in luminosity in Figures 1-4 will be an overestimate. In Figure 7, we plot the exact solution for magnetic dipole field strength versus $T_{\rm sph}$ versus $\Delta L/L_{\rm disc, o}$ for the limiting case of very mild beaming with $b$ = 0.5 (corresponding to a scale height of only around unity). As can be seen by comparing to panel B of Figures 1-4, the greatest impact occurs where advection of energy is included, permitting smaller values of $\Delta L/L_{\rm disc, o}$ to be reached for a given dipole field strength, as expected. The impact at low $\epsilon_{\rm wind}$ and without advection is minimal, which does not affect the constraints we have placed on the dipole field strength in NGC 1313 X-1 (modulo the caveats already discussed).

In future we hope to build on this initial work by modelling the complete ULX spectrum as it transits into the propeller phase, incorporating the anisotropy of the disc and the key temperatures of the various components  (\citealt{Poutanen_2007_ln, Mushtukov_2017_opt_thick_env_NS}).

\section*{Acknowledgements}

The authors thank the anonymous referee for their useful suggestions. MM and AG acknowledge support via STFC Consolidated grant (ST/V001000/1). The authors thank Hugh Dickinson for useful advice.  
\section*{Data Availability}

Data can be made accessible upon request. 
 


\bibliographystyle{mnras.bst}
\bibliography{bibliography.bib}  





\bsp	
\label{lastpage}
\end{document}